\documentclass[sigconf]{acmart}
\usepackage{flushend,cuted}
\usepackage{enumitem}
\usepackage{booktabs}
\usepackage{graphicx}
\usepackage{array}
\usepackage{lineno}
\usepackage{multirow}
\usepackage{hhline}
\usepackage{makecell}
\setlist[itemize]{leftmargin=10pt}
\usepackage[marginal]{footmisc}

\AtBeginDocument{%
  \providecommand\BibTeX{{%
    \normalfont B\kern-0.5em{\scshape i\kern-0.25em b}\kern-0.8em\TeX}}}

\copyrightyear{2025}
\acmYear{2025}
\setcopyright{acmlicensed}
\acmConference[WWW Companion '25]{Companion Proceedings of the ACM Web Conference 2025}{April 28-May 2, 2025}{Sydney, NSW, Australia}
\acmBooktitle{Companion Proceedings of the ACM Web Conference 2025 (WWW Companion '25), April 28-May 2, 2025, Sydney, NSW, Australia}
\acmDOI{10.1145/3701716.3715251}
\acmISBN{979-8-4007-1331-6/25/04}

\begin{document}

\title{NLGR: Utilizing Neighbor Lists for Generative Rerank in Personalized Recommendation Systems}


\author{Shuli Wang}
\authornote{Corresponding author.}
\affiliation{%
  \institution{Meituan}
   \city{Chengdu}
  \country{China}
}
\email{wangshuli03@meituan.com}

\author{Xue Wei}
\affiliation{%
\institution{Meituan}
   \city{Chengdu}
  \country{China}
  }
\email{weixue06@meituan.com}

\author{Senjie Kou}
\affiliation{%
\institution{Meituan}
   \city{Chengdu}
  \country{China}
  }
\email{kousenjie@meituan.com}

\author{Chi Wang}
\affiliation{
\institution{Meituan}
   \city{Chengdu}
  \country{China}
  }
\email{wangchi06@meituan.com}

\author{Wenshuai Chen}
\affiliation{%
\institution{Meituan}
   \city{Chengdu}
  \country{China}
  }
\email{chenwenshuai@meituan.com}

\author{Qi Tang}
\affiliation{%
  \institution{Meituan}
   \city{Chengdu}
  \country{China}
}
\email{tangqi22@meituan.com}

\author{Yinhua Zhu}
\affiliation{%
\institution{Meituan}
   \city{Chengdu}
  \country{China}
  }
\email{zhuyinhua@meituan.com}

\author{Xiong Xiao}
\affiliation{%
\institution{Meituan}
   \city{Chengdu}
  \country{China}
  }
\email{xiaoxiong02@meituan.com}

\author{Xingxing Wang}
\affiliation{%
\institution{Meituan}
   \city{Beijing}
  \country{China}
  }
\email{wangxingxing04@meituan.com}


\renewcommand{\shortauthors}{Shuli Wang et al.}

\begin{abstract}
Reranking plays a crucial role in modern multi-stage recommender systems by rearranging the initial ranking list. Due to the inherent challenges of combinatorial search spaces, some current research adopts an evaluator-generator paradigm, with a generator generating feasible sequences and an evaluator selecting the best sequence based on the estimated list utility. However, these methods still face two issues. Firstly, due to the goal inconsistency problem between the evaluator and generator, the generator tends to fit the local optimal solution of exposure distribution rather than combinatorial space optimization. Secondly, the strategy of generating target items one by one is difficult to achieve optimality because it ignores the information of subsequent items.

To address these issues, we propose a utilizing \textbf{N}eighbor \textbf{L}ists model for \textbf{G}enerative \textbf{R}eranking (NLGR), which aims to improve the performance of the generator in the combinatorial space.
NLGR follows the evaluator-generator paradigm and improves the generator's training and generating methods. 
Specifically, we use neighbor lists in combination space to enhance the training process, making the generator perceive the relative scores and find the optimization direction.
Furthermore, we propose a novel sampling-based non-autoregressive generation method, which allows the generator to jump flexibly from the current list to any neighbor list.
Extensive experiments on public and industrial datasets validate NLGR's effectiveness and we have successfully deployed NLGR on the Meituan food delivery platform.
\end{abstract}





\maketitle

\section{Introduction}
E-commerce platforms, such as Meituan and Taobao, need to provide users with personalized services from millions of items. To improve recommendation efficiency, personalized recommendation systems generally include three stages: matching,  ranking, and reranking. The ranking models (e.g.,Wide\&Deep \cite{W&D}, DeepFM \cite{deepfm}, DIN \cite{din}) evaluate the point-wise items respectively based on the Click-Through Rate (CTR), but they ignore the crucial mutual influence among items \citeN{chang2023twin,pi2020search}. 
Research \citeN{burges2010ranknet, listnet, ai2018learning, pang2020setrank} indicates that optimizing a list-wise utility is a more advantageous strategy, as it capitalizes on the mutual influences between items within the list to enhance overall performance.

The key challenge of reranking is to explore the optimal list in the huge combinatorial space \citeN{gfn,nar4rec}. Initially, some list-wise methods \citeN{prm,midnn,dlcm} re-evaluate and score items within lists by modeling the context. These list-wise methods can obtain more accurate scores than point-wise methods, then they use a greedy strategy to reorder based on the list-wise score. However, these methods face the evaluation-before-reranking problem \citeN{xi2021context, grn} and cannot achieve optimization in combinatorial space.
To resolve the problem, a straightforward solution is to evaluate every possible permutation, which is global-optimal but is too complex to meet the strict inference time constraint in industrial systems. Therefore, most existing evaluator-based reranking framework uses a two-stage architecture \citeN{feng2021revisit, xi2021context, grn, nar4rec, dcdr} which consists of a generator and an evaluator.  Within the generator-evaluator paradigm, the generator plays a crucial role \cite{nar4rec}. Some methods use heuristic methods as generators, such as beam-search \cite{medress1977speech} and SimHash \cite{chen2021end}. The generators of these methods do not utilize the information of the evaluator, resulting in limited effectiveness. Recently, some methods \citeN{grn,nar4rec,dcdr} utilize generative models as generators and achieve better results than heuristic methods.

However, existing generative reranking methods face two significant challenges. Firstly, due to the goal inconsistency problem, the generator has difficulty finding the optimal list in the combinatorial space. While the evaluator is trained to fit list-wise scores of items, the generator is tasked with transforming any candidate list into the optimal one. This disparity in objectives between the evaluator and the generator complicates the transfer of guidance, often causing the generator to merely fit exposure distributions in extreme cases. Secondly, the strategy of generating target items sequentially, one by one, hinders the achievement of optimal results. The sequential decoding process focuses solely on preceding items, neglecting information from succeeding items. This limitation leads to suboptimal performance as the model fails to fully leverage the available context.

\begin{figure}[h]
\centering
\includegraphics[width=\linewidth, height=2\textheight, keepaspectratio]{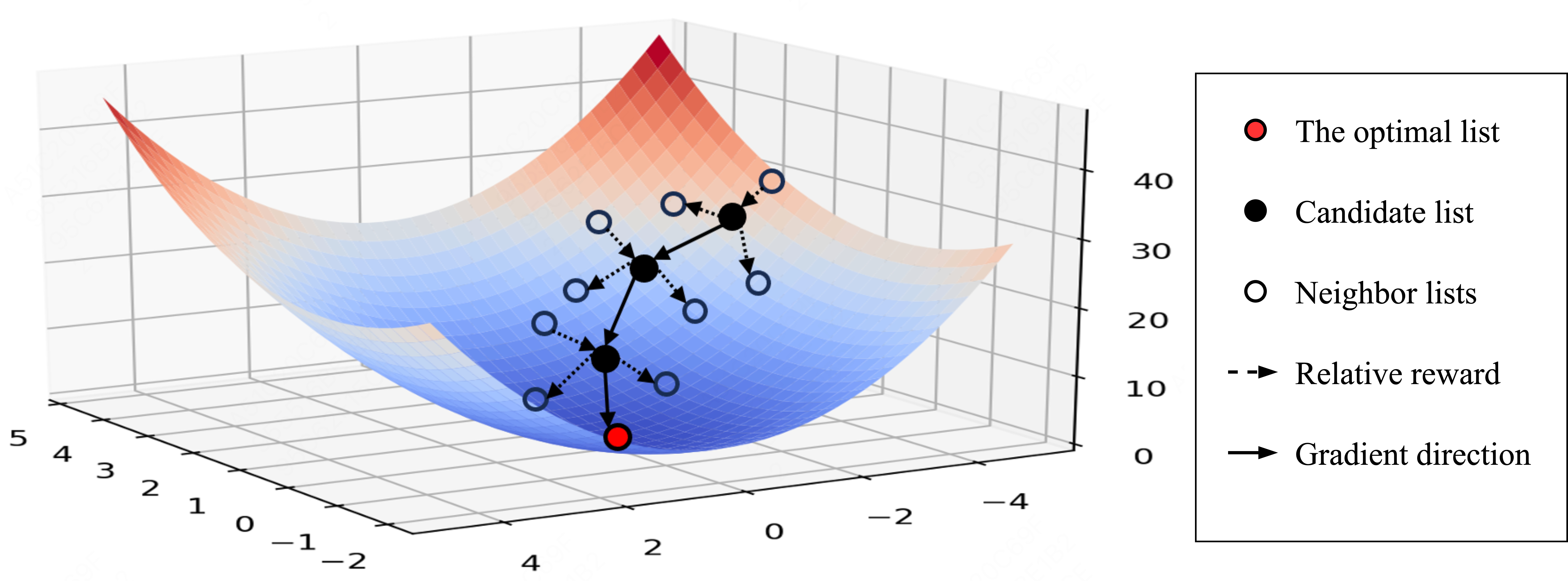}

\caption{Generator optimization legend: reach the optimal step by step under the guidance of the neighbor lists.}
\label{fig:gradient}
\end{figure}

To address the aforementioned challenges, we propose a novel \textbf{G}enerative \textbf{R}eranking method that utilizes \textbf{N}eighbor \textbf{L}ists, named NLGR. Our approach still follows the evaluator-generator paradigm, with the evaluator solely assisting in the generator's offline training. Our improvements are twofold as follows. First, we introduce an enhanced training process that utilizes neighbor lists, enabling the generator to perceive relative scores within the combinatorial space and identify the optimal direction. As depicted in Figure \ref{fig:gradient},  by evaluating multiple neighboring lists and iterating several times, the generator will converge to an optimal state. Second, we propose a sampling-based non-autoregressive generation method. This method first determines the position of the item that needs to be replaced using the Position Decision Unit (PDU), and then retrieves new replacement items from the candidate item set using the Candidate Retrieval Unit (CRU), which allows for a more flexible exploration of the combinatorial space to find the optimal list.

In summary, the contributions of this work are as follows:
\begin{itemize}
\item We propose a novel generative reranking method that utilizes neighbor lists to address the goal inconsistency problem between evaluators and generators. To the best of our knowledge, we are the first to propose and attempt to solve this problem.

\item We propose a novel sampling-based non-autoregressive generation method that generates the optimal list more flexibly in the combination space.

\item We have verified the superior performance of NLGR through extensive experiments on both offline and online datasets. It is notable that NLGR has been deployed in the Meituan food delivery platform and has achieved significant improvement under various metrics.

\end{itemize}

\section{Related Work}
\subsection{Reranking Methods}

Typical reranking methods can be divided into two categories \cite{pier}. The first category is the one-stage reranking methods, which only generates one list as output by capturing the mutual influence among items. Seq2slate \cite{seq2slate} utilizes pointer-network and MIRNN \cite{zhuang2018globally} utilizes GRU to determine the item order one-by-one. Methods such as PRM \cite{prm} and DLCM \cite{ai2018learning} take the initial ranking list as input, use RNN or self-attention to model the context-wise signal, and output the predicted value of each item. Such methods bring an evaluation-before-reranking problem \cite{xi2021context} and lead to suboptimal. Similarly, methods such as EXTR \cite{extr} estimate pCTR of each candidate item on each candidate position, which are substantially point-wise models and thus limited in extracting exact context. MIR \cite{mir} capturing the set2list interactions by a permutation-equivariant module
Another category is the two-stage reranking methods, which tries to evaluate every possible permutation through a well-designed context-wise model. This is a global-optimal method but is too complex to meet the strict inference time constraint in industrial systems. To reduce the complexity, PRS \cite{feng2021revisit} adopts beam-search to generate a few candidate permutations first, and score each permutation through a permutation-wise ranking model. PIER \cite{pier} applies SimHash to select top-K candidates from the full permutation.  

\subsection{ Generative Reranking Solutions}

In recent years, the generative reranking model \citeN{listcvae, pivotcvae, slateq} for listwise recommendation has been a topic of discussion. To manage the vast combinatorial output space of lists, the generative approach directly models the distribution of recommended lists and employs deep generative models to generate a list as a whole. For instance, ListCVAE \cite{listcvae} utilizes conditional variational autoencoders (CVAE) to capture the positional biases of items and the interdependencies within the list distribution. But Pivot-CVAE \cite{pivotcvae} indicates that ListCVAE suffers from a trade-off dilemma between accuracy and diversity, and proposes an "elbow" performance to enhance the accuracy-based evaluation.

GFN \cite{gfn} uses a flow-matching paradigm that maps the list generation probability with its utility. Essentially it is still studying list distributions rather than directly modeling the permutation space, so it still has the challenge mentioned above. GRN \cite{grn} proposes an evaluator-generator framework to replace the greedy strategy, but it can't avoid the evaluation-before-reranking problem \cite{xi2021context} because it takes the rank list as input to the generator. DCDR \cite{dcdr} introduces diffusion models into the reranking stage and presents a discrete conditional diffusion reranking framework. NAR4Rec \cite{nar4rec} uses a non-autoregressive generative model to speed up sequence generation. However, these methods still face the two problems mentioned above.

\section{Problem Definition}
In the Meituan food delivery platform, we adhere to the matching, ranking, and reranking recommendation paradigms to present items to users in a list format.
Initially, we define a user set $U$ and an item set $I$. We utilize the session-level users' historical interacted lists $B$ and candidate set $C$ to represent the user's features $u \in U$, which consistently holds in our reranking scenario $C \in I$, and ultimately select the list $L$ for users. 

Reranking introduces a combination space with exponential size, represented as $\mathcal{O}(A_n^m)$, where $n$ represents the size of the candidate set $C$ and $m$ represents the size of the output list $L$. 
Our optimization objective is to learn a strategy $\pi: C \rightarrow L$ by maximizing the list score reward $R(u,\pi)$. The list score reward $R(u, \pi)$ takes into account factors such as click-through rate (CTR) and conversion rate (CVR). 

\textbf{Neighbor List.} If the distance between two lists is 1, that is, the two lists only differ by 1 item, we define the two lists as neighbor lists. If we swap two items in a list, the new list will have a distance of 2.

\section{Proposed Method}

In this section, we present a detailed introduction of NLGR. 
We first introduce the evaluator (in Section \ref{section:evaluator}) and generator (in Section \ref{section:generator}) of NLGR, denoted as NLGR-E and NLGR-G respectively. Then, we demonstrate the offline training process of NLGR in Section \ref{section:offline_training}. The evaluator is only assisting in the offline training of the generator, so NLGR is a fast inference method.

\subsection{Evaluator Model}\label{section:evaluator}

We use NLGR-E to evaluate a ranked sequence, such as the exposed list which is displayed to the user, or the candidate lists generated by NLGR-G.
The structure of NLGR-E is shown on the left side of Figure \ref{fig:NLGR}. 

NLGR-E includes two inputs: the exposed list to be evaluated and the user session-level behavior sequence, where each session in the user session-level behavior sequence is the user's historical exposed list. The evaluation process of NLGR-E is as follows: 

First, we use an embedding layer to get the embedding of the original input, donated as $\mathbf{X} \in \mathbb{R}^{m \times F \times D}$ and $\mathbf{M} \in \mathbb{R}^{H \times m \times F \times D}$ respectively, where $H$ represents the number of history sessions, $m$ represents the number of items in each list, $F$ represents the number of feature fields for each item (i.g., ID, category, position index), and $D$ represents the dimension of the embedding. 
Inspired by DIF \cite{dif}, to avoid feature interference, we propose the D-Attention unit to \underline{d}ecouple the feature context information.
We first calculate the attention score on $i$-th attribute $\mathbf{X}_{i} \in \mathbb{R}^{m \times D}$:
\begin{equation} \label{equation:1}
\mathrm{Att}_{i} = \sigma \left( \frac{(\mathbf{X}_{i} \mathbf{W}^Q_{i})(\mathbf{X}_{i} \mathbf{W}^K_{i})^\top}{\sqrt{D}}\right),   \forall i \in [F] , 
\end{equation}
where $\mathbf{W}_{i}^Q$ and $\mathbf{W}_{i}^K \in \mathbb{R}^{D \times D}$ denote the weight matrices.

Then we aggregate all attention matrices $\mathrm{Att}_{i} \in \mathbb{R}^{m \times m}$ and get the item-level attention score $\mathrm{Att_{all}} \in \mathbb{R}^{m \times m}$:
\begin{equation} \label{equation:2}
\mathrm{Att_{all}} = \frac{1}{F} \sum_{f=1}^{F} \mathrm{Att}_{i}.
\end{equation}

Subsequently, we aggregate ID feature embeding $\mathbf{X}^{id} \in \mathbb{R}^{m \times D}$ and obtain each exposed list's representation $\mathbf{e}^l \in \mathbb{R}^{D}$:
\begin{equation} \label{equation:3}
\mathbf{e}^l = \mathrm{reduce\_mean}\left(\mathrm{Att_{all}}(\mathbf{X}^{id}\mathbf{W}^V)\right).
\end{equation}

Similarly, by performing the above operations on each session in the user session-level behavior sequence, we can obtain the representation $\mathbf{e}^s_i \in \mathbb{R}^{ D} $ of each session. Then we input $\mathbf{e}^s_i$ into a Self-Attention layer (SA) \citeN{vaswani2017attention,kang2018self} to get the user representation $\mathbf{e}^u \in \mathbb{R}^{D}$:
\begin{equation} \label{equation:4}
\mathbf{e}^u = \mathrm{SA}(\mathbf{e}^s_1 || \mathbf{e}^s_2 ||...||\mathbf{e}^s_H),
\end{equation}
where || represents concatenate operate. 

Finally, the predicted Click-Through Rate (pCTR) of the $j$-th item can be represented as:
\begin{equation} \label{equation:5}
\hat{y}_j=\sigma(\mathrm{Tiled\_MLP}(\mathbf{X}_j || \mathbf{e}^l || \mathbf{e}^u || PE_j)),
\end{equation}
where $PE_j$ denotes $j$-th position embedding. Similarly, the pCVR and other evaluations also follow the above process. 

\subsection{Generator model} \label{section:generator}
\begin{figure*}[h]
\centering
\includegraphics[width=\textwidth]{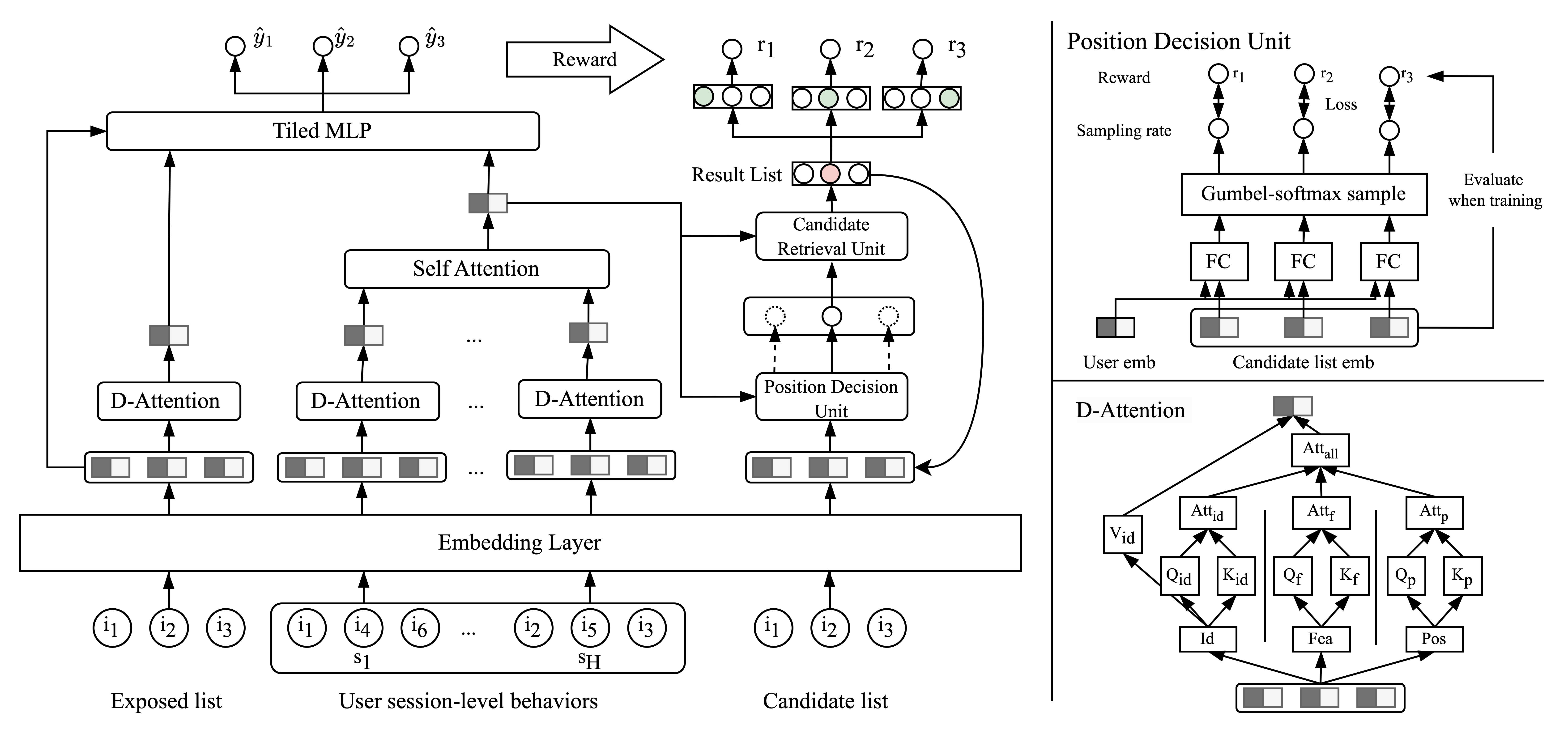}
\caption{The overall architecture of NLGR.}
\label{fig:NLGR}
\end{figure*}

We utilize NLGR-G to generate the optimal list in combinatorial space. The structure of NLGR-G is shown on the middle side of Figure \ref{fig:NLGR}. 

Similar to NLGR-E, NLGR-G includes two inputs: the candidate list to be reranked and the user session-level behavior sequence, where each session in the user session-level behavior sequence is the user's historical exposed list. Theoretically, the candidate list can be any in the combinatorial space, but generally, we use the ranking list as the initial input. The generation process of NLGR-G is as follows:

First, we obtain the user representation $\mathbf{e}^u$ through Eq. \ref{equation:4}. These parameters are shared from NLGR-E to ensure that the remaining parameters can be optimized more focused.

Then, we propose a sampling-based non-autoregressive generation method. It first determines the position of the item that needs to be replaced through the Position Decision Unit (PDU), then retrieves new replacement items from the candidate item set through the Candidate Retrieval Unit (CRU).

\subsubsection{Position Decision Unit (PDU)}
First, we use the embedding layer to get the embedding of the candidate list, donated as $\mathbf{X} \in \mathbb{R}^{m \times F \times D}$. Then, we flatten $j$-th item embedding $\mathbf{X}_j \in \mathbb{R}^{F \times D}$ and use a Fully Connected layer (FC) to calculate the selected logit of the $j$-th position:
\begin{equation} \label{equation:6}
h_j=\mathrm{FC}_1(\mathbf{X}_j || \mathbf{e}^l || \mathbf{e}^u || PE_j).
\end{equation}

To solve the non-differentiable problem of the sampling distribution, inspired by \citeN{jang2016categorical,liu2021neural,huijben2022review}, we use the Gumbel-softmax trick for sampling:
\begin{equation} \label{equation:7}
r^p_j=\mathrm{softmax}\left(\frac{log(h_j)+n}{\tau}\right),\forall j \in [m],
\end{equation}
where $\tau > 0$ is a temperature parameter, $n=-\mathrm{log}(-\mathrm{log}(u)))$ represents random noise sampled from the Gumbel distribution, $u$ is a uniform distribution between [0, 1]. During backpropagation, the gradient is calculated using Eq. \ref{equation:7}. While during forward propagation, the replaced position is $j=\mathrm{argmax}(r^p_j)$.
 
\subsubsection{Candidate Retrieval Unit (CRU)} 
After determining the position $j$ to be replaced, we need to select a suitable one from $n$ candidate items and place it at position $j$. Since this operation will be repeated multiple times during the generation process, we propose leveraging retrieval-based techniques to quickly achieve this goal for efficiency. 
First, we mask the position $j$ of the candidate list, denoted as $\mathbf{X}_j^{mask} \in \mathbb{R}^{m \times F \times D}$, and then extract the list representation $\mathbf{e}_j^{mask} \in \mathbb{R}^{D}$:
\begin{equation} \label{equation:8}
\mathbf{e}_j^{mask}=\mathrm{SA}(\mathbf{X}_j^{mask}).
\end{equation}

We then compute the representation of each candidate item at position $j$:
\begin{equation} \label{equation:9}
\mathbf{e}^c_k=\mathrm{FC}_2(\mathrm{flatten}(\mathbf{X}^c_k) || PE_j), \forall k \in [n],
\end{equation}
where $\mathbf{X}^c \in \mathbb{R}^{n \times F \times D}$ denotes embedding of candidate set $C$, and $\mathbf{X}^c_k$ denotes $k$-th candidate item's embedding.

Then we use an FC layer to calculate the selected logit of the $j=k$-th candidate item:
\begin{equation} \label{equation:10}
g_k=\mathrm{FC}_3(\mathbf{e}^c_k || \mathbf{e}_j^{mask} || \mathbf{e}^u || PE_j).
\end{equation}

Similarly, to overcome non-differentiable problems, we use the Gumbel-softmax trick for sampling:
\begin{equation} \label{equation:11}
r^c_k=\mathrm{softmax}\left(\frac{log(g_k)+n}{\tau}\right),\forall k \in [n],
\end{equation}
During backpropagation, the gradient is calculated using Eq. \ref{equation:10}. While during forward propagation, the newly inserted item is $c=\mathrm{argmax}(r^c_k)$.

\textbf{Stop condition.} Note that the generation process may be repeated many times until the newly inserted item equals the replaced item or the values of $r^p_j$ and $r^c_k$ are too low.

\subsection{Utilizing Neighbor Lists Training} \label{section:offline_training}
In this section, we elaborate on the offline training process of NLGR, which includes the training procedures for NLGR-E and NLGR-G. As mentioned before, the evaluator is trained to fit list-wise scores of items, and the generator is tasked with transforming any ranking list into the optimal one. This goal inconsistency between the evaluator and the generator complicates the transfer of guidance. We will introduce our solution in detail below.

\subsubsection{\textbf{Training of NLGR-E}}
To accurately evaluate the return of the exposure list and estimate the listwise pCTR value of the exposure list, we train NLGR-E using real data collected from online logs. The input is the features of the recommended advertisement sequences exposed in reality online, and the advertising return situation, including exposure, click, conversion, and other performance indicators, is used as the label to supervise the training of NLGR-E, enabling it to accurately evaluate the return of the recommended sequence. The loss of NLGR-E is calculated as follows:

\begin{equation}
\mathcal{L}^E  =\sum_{j=1}^m \left( - y_{j}  \log(\widehat {y}_{j})-(1- y_{j})  \log  {(1-\widehat {y}_{j}} ) \right),
\end{equation}
where $y_j$ represents the real label, $\widehat{y}_{j}$ represents the predicted value of NLGR-E, and the evaluation is carried out for the $m$ items in the exposure list in turn.

\subsubsection{\textbf{Training of NLGR-G}} \label{section:train_G}

To address the problem of goal inconsistency mentioned before, we use neighbor lists to guide NLGR-G within the counterfactual space. For each list generated by NLGR-G, NLGR-E simulates human feedback and provides a reward $R$ to guide NLGR-G training. 

\begin{figure}[h]
\centering
\includegraphics[width=0.45\textwidth]{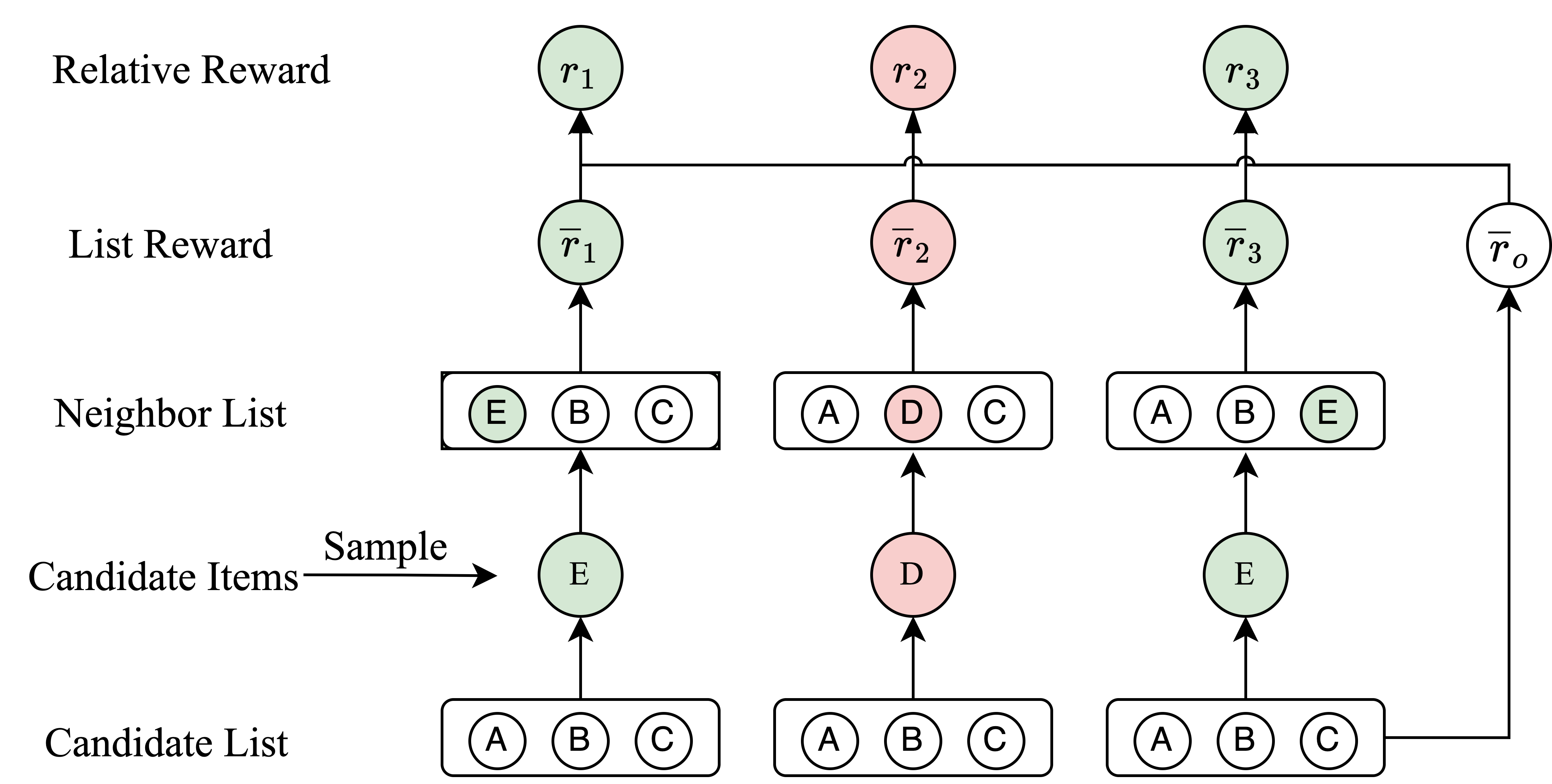}
\caption{The training process of NLGR-G for a candidate list of length 3.}
\label{fig: counterfactual}
\end{figure}

Figure \ref{fig: counterfactual} shows an example of the NLGR-G training process. First, for each candidate list $L^o=[i^o_1,i^o_2,...,i^o_m]$, we sample a replacement item $i^*$ from the candidate set $C=[i_1,i_2,...,i_n]$, replace the $j$-th item $L_o$'s, and construct a neighbor list $L^*_j=[i^o_1,i^o_2,...i^ *_j,...,i^o_m]$. Repeating and replacing each position, we can get a set of neighbor lists $L^*=[L^*_1,L^*_2,...,L^*_m]$.

Then, we use the trained NLGR-E to evaluate the candidate list $L^o$ and the neighbor list $L^*$, and indicators such as pCTR and pCVR will be estimated. We convert the estimated value into list reward based on business indicators as follows:
\begin{equation}\label{equation:r}
\overline{r}=
\begin{cases}
e^{w-1} -1 & \text{if }w>1 \\
0          & \text{if }w=1 \\
1-e^{1-w}  & \text{if }w<1
\end{cases},
\end{equation}

\begin{equation}
w = k1\cdot L_{ctr}+k2\cdot L_{ctr}\cdot L_{cvr},
\end{equation}
where $L_{ctr}$ and $L_{cvr}$ represent the list total pCTR and pCVR which are evaluated by NLGC-E, respectively. The parameters $k_1$ and $k_2$ are business parameters that depend on the click bid and conversion price in the specific business.

Through Eq. \ref{equation:r}, we can get the rewards of the neighbor lists $L^*$ and the original candidate list $L^o$, denoted as $\overline{r}=[\overline{r}_1, \overline{r}_2,...,\overline{r}_m]$ and $\overline{r_o}$ respectively. NLGR-G is tasked to iterate from the candidate list to a more optimal list, so we calculate the relative reward for each position $j$:
\begin{equation} \label{equation:R}
r_j = \overline{r}_j - \overline{r}_o, \forall j \in [m].
\end{equation}

The authentic evaluation $R$ for the candidate list $L^o$ is obtained by aggregating the relative rewards of all positions. And we define the counterfactual loss of NLGR-G as $-R$:
\begin{equation} \label{equation:R}
\mathcal{L}^G_ {1} = -R = -\sum_{j=1}^{m} r_j.
\end{equation}

Furthermore, to increase the stability of the NLGR-G's generation process, we propose using the position reward $r_j$ to provide additional guidance to the Position Decision Unit (PDU). Specifically, we introduce cross-entropy loss as an auxiliary loss to measure the distribution difference of position sampling $r^p$ and position reward $r_j$:

\begin{equation}
\mathcal{L}^G_{2}  = -\sum_{j=1}^m \mathrm{Norm}(r_j) \cdot \mathrm{log}r^p_j,
\end{equation}
where $\mathrm{Norm}(r_j)=\frac{r_j}{\sum r_j}$.

The final loss of NLGR-G in each batch is: 
\begin{equation}\label{equation:loss_G}
\mathcal{L}^G =\frac{1}{|B|}\sum_{B}
(\mathcal{L}_1^G + \alpha \cdot \mathcal{L}_2^G),
\end{equation}
where $\alpha$ is a coefficient to balance the two losses.
 
\section{Experiments}
To validate the superior performance of NLGR, we conducted extensive offline experiments on the Meituan dataset and verified the superiority of NLGR in online A/B tests. In this section, we first introduce the experimental setup, including the dataset and baseline. Then, in Section \ref{exp_result}, we present the results and analysis of various reranking methods in both offline and online A/B tests.

\subsection{Experimental Setup}
\subsubsection{Dataset}
In order to verify the effectiveness of NLGR, we conduct sufficient experiments on both public dataset and industrial dataset. For public dataset, we choose Taobao Ad dataset\footnote{https://tianchi.aliyun.com/dataset/56}. For the industrial dataset, we use real-world data collected from Meituan food delivery platform. 

\begin{itemize}[leftmargin=*]
\item  Taobao Ad. It is a public dataset collected from the display advertising system of Taobao. This dataset contains more than 26 million interaction records of 1.14 million users within 8 days. Each sample comprises five features: user ID, timestamp, behavior type, item brand ID, and category ID. 
\item Meituan. It is an industrial dataset collected from the Meituan food delivery platform during October 2023, which contains 1.3 billion interaction records of 130 million users within 30 days. The dataset includes 239 features, two labels: click and conversion, and collects all items on the same page as one record. We divide the dataset into training and test sets with a proportion of 9:1.
\end{itemize}

Table \ref{tab:my_table} gives a brief introduction to the datasets.
\begin{table}[H]
\caption{Statistics of datasets}
\label{tab:my_table}
\begin{tabular}{cccc}
\hline
\textbf{Dataset} & \textbf{\#Users} & \textbf{\#Items} & \textbf{\#Records} \\ \hline
Taobao Ad               & 1,141,729        & 99,815           & 26,557,961           \\
Meituan          & 130,648,310      & 14,054,691       & 1,331,247,488        \\ \hline
\end{tabular}
\end{table}

\begin{table*}[htb]
\caption{Performance comparison. The best result and the second-best result in each column are in bold and underlined}
\begin{tabular}{llccccccccc}
\toprule[1pt]
\multirow{2}{*}{Dataset} & \multirow{2}{*}{Metric} & \multicolumn{2}{c}{Group I} 
& \multicolumn{1}{c}{} 
& \multicolumn{2}{c}{Group II} 
& \multicolumn{1}{c}{} 
& \multicolumn{2}{c}{Group III} & \multirow{2}{*}{NLGR} \\
    \cline{3-4} 
    \cline{6-7}
    \cline{9-10} 
 & & DNN & DeepFM & & PRM & MIR & & GRN & DCDR &  \\ \hline
\multirow{5}{*}{Taobao Ad} 
&AUC             & 0.5869 & 0.5871 & & 0.6052 & 0.6047 & & 0.6101 & \underline{0.6217} & \textbf{0.6344} $\pm$ 0.5\textperthousand\\
&LogLoss         & 0.1878 & 0.1866 & & 0.1842 & 0.1853 & & 0.1820 & \underline{0.1792} & \textbf{0.1749} $\pm$ 0.1\textperthousand\\
&NDCG@10         & 0.1527 & 0.1548 & & 0.1805 & 0.1769 & & 0.1896 & \underline{0.2038} & \textbf{0.2323} $\pm$ 0.2\textperthousand\\
&NDCG@5          & 0.1092 & 0.1110 & & 0.1206 & 0.1193 & & 0.1273 & \underline{0.1453} & \textbf{0.1830} $\pm$ 0.1\textperthousand\\
\midrule
\multirow{5}{*}{Meituan} 
&AUC             & 0.7034 & 0.7070 & & 0.8096 & 0.8031 & & 0.8034 & \underline{0.8181} & \textbf{0.8349} $\pm$ 0.4\textperthousand \\
&LogLoss         & 0.1162 & 0.1162 & & 0.1096 & 0.1102 & & 0.1102 & \underline{0.1087} & \textbf{0.1039} $\pm$ 0.1\textperthousand \\
&NDCG@10         & 0.2015 & 0.2019 & & 0.2743 & 0.2742 & & 0.2744 & \underline{0.2793} & \textbf{0.2857} $\pm$ 0.2\textperthousand \\
&NDCG@5          & 0.1569 & 0.1580 & & 0.2378 & 0.2365 & & 0.2353 & \underline{0.2400} & \textbf{0.2431} $\pm$ 0.2\textperthousand\\
\bottomrule[1pt]
\end{tabular}

\label{tab:tab2}
\end{table*}

\begin{table}[h]
\caption{Hit ratio comparison. The best result and the second-best result in each column are in bold and underlined}
\label{tab:3}
\begin{tabular}{l|cc|cc}
\toprule
\multirow{2}{*}{Model} & \multicolumn{2}{c|}{Taobao Ad} & \multicolumn{2}{c}{Meituan}  \\
 & HR@10\% & HR@1\%   & HR@10\% & HR@1\% \\
\midrule
PRM           & 0.1125             & 0.0794             & 0.5702             & 0.4412  \\
GRN           & \underline{0.2160} & 0.0829             & 0.7386             & 0.5783  \\
DCDR          & 0.2159             & \underline{0.0835} & \underline{0.7573} & \underline{0.5802}\\
NLGR          & \textbf{0.4091}    & \textbf{0.3220}    & \textbf{0.8369}    &\textbf{0.7523} \\
\bottomrule
\end{tabular}
\end{table}

\subsubsection{Baseline}
The following six baselines are chosen for comparative experiments and divided into three groups. We select DNN and DeepFM as point-wise baselines (Group I), PRM and MIR as list-wise baselines (Group II), and GRN and DCDR as generative baselines (Group III). A brief introduction of these methods is as follows:

\begin{itemize}[leftmargin=*]
\item $\textbf{DNN}$\cite{dnn} is a basic deep learning method for CTR prediction, which applies MLP for high-order feature interaction.
\item $\textbf{DeepFM}$\cite{deepfm} is a general deep model for recommendation, which combines a factorization machine component and a deep neural network component.
\item $\textbf{PRM}$\cite{prm} adjusts an initial list by applying the self-attention mechanism to capture the mutual influence between items.
\item $\textbf{MIR}$\cite{mir} learns permutation-equivariant representations for the inputted items via self-attention.
mechanism to capture the mutual influence between items.
\item $\textbf{GRN}$\cite{grn} is a generative reranking model which consists of the evaluator for predicting interaction probabilities and the generator for generating reranking results.
\item $\textbf{DCDR}$\cite{dcdr} presents a discrete conditional diffusion reranking framework.
\end{itemize}

\subsubsection{Evaluation Metrics. }
We adopt several metrics, i.e., \textbf{AUC} (Area Under ROC Curve), \textbf{Logloss} and \textbf{NDCG} (normalized discounted cumulative gain) to evaluate NLGR-E in offline experiments. A larger AUC and NDCG indicate better recommendation performance, while Logloss performs the opposite. 

We use \textbf{HR} (Hit Ratio) \cite{alsini2020hit} to evaluate NLGR-G in offline experiments. It is worth noting that only one list produced by reranking algorithms can be presented to the user. As a result, the generator cannot be fully and fairly evaluated. A practical workaround is to employ the evaluator to assess the performance of the generator. For each data, we evaluate all candidate lists using NLGR-E. HR@10\% is 1 only when the rerank list produced by NLGR-G is ranked within the top 10\% as sorted by NLGR-E. The HR metrics are only meaningful with evaluator-based reranking methods. The results of HR can be seen in Table \ref{tab:3}.

It is worth noting that AUC and HR can measure the evaluator and generator respectively. AUC measures the model's ability to evaluate an ordered list, while HR measures the consistency between the evaluator and generator. A decrease in any indicator will reduce the recommendation effect.

In online experiments, we adopt CTR and GMV (Gross Merchandise Volume) as evaluation metrics.

\subsubsection{Implementation Details}
We implement all the deep learning baselines and NLGR with TensorFlow 1.15.0 using NVIDIA A100-SXM4-80GB. For all comparison models and our NLGR model, we adopt Adam as the optimizer with the learning rate fixed to 0.001 and initialize the model parameters with normal distribution by setting the mean and standard deviation to 0 and 0.01, respectively. The batch size is 1024, the embedding size is 8, the $\alpha$ is 0.2. The hidden layer sizes of tiled MLP are (1024, 256, 128). 
For the Taobao Ad dataset, the length of the ranking list and reranking list are both 5, thus the length of full permutation is 120. For Metuan dataset, we select 4 items from the initial ranking list which contains 12 items, thus the length of full permutation is $A_{12}^4=11,880$. 
All experiments are repeated five times and the averaged results are reported.

\subsection{Experimental Results}\label{exp_result}
\subsubsection{Performance Comparison}

Table \ref{tab:tab2} and Table \ref{tab:tab3} summarize the results of offline experiments. We have the following observations from the experimental results: 
\begin{enumerate}[leftmargin=*]
\item [i)] PRM in Group II outperforms all models in Group I, which verifies the impact of the mutual influence among contextual items. DCDR in Group III outperforms all the other models in Groups I and II, which verifies the effectiveness of generative methods.
\item [ii)] DCDR indeed exhibits robust generative capabilities, thanks to the incorporation of the diffusion model, and achieves the second-best result. Nevertheless, DCDR overlooks the significance of full sight and falls short of leveraging the evaluator's potential to its fullest, which limits its effect. 
Our proposed NLGR significantly and consistently outperforms the state-of-the-art approaches in all metrics on both datasets. As presented in Table \ref{tab:tab2}, our proposed NLGR brings 0.8349/0.6344 absolute AUC, 0.2857/0.2323 absolute NDCG@10, 0.2431/0.1830 absolute NDCG@5 on Metuan/Ad dataset, gains significant improvement in industrial recommendation system. NLGR has greater improvements on Meituan dataset because Meituan dataset has more realistic reranking scenarios and richer features.
\item [iii)] Our proposed NLGR brings 0.8369/0.4091 absolute HR@10\% on Metuan/Ad dataset. This demonstrates our generator achieves extreme improvements via counterfactual evaluation. Compared with GRN and PRM, there is a huge improvement. As a typical greedy reranking algorithm, PRM only has 0.5702/0.1125 absolute HR@10\% on Metuan/Ad dataset. This conclusively shows that effective rearrangement cannot be attained by relying solely on a greedy strategy. As a generative reranking model, DCDR lacks full sight, which is why its HR@10\% is not optimal.
\end{enumerate}




\subsubsection{Ablation Study}
To verify the impact of different units, we study three ablated variants of NLGR on Meituan dataset.
\begin{itemize}[leftmargin=*]
\item NLGR without relative reward $r$. To verify the effectiveness of the neighbor list training method in addressing the goal inconsistency problem, this variant removes the relative reward defined in Eq. \ref{equation:R} and replaces it with the predicted value returned by the evaluator directly.
\item NLGR with autoregressive generation (AG). To verify the sampling-based non-autoregressive generation method in NLGR-G, we replaced it with pointer network \cite{vinyals2015pointer}, which is a sequence generation method widely used in reranking models \citeN{grn,seq2slate}.
\item NLGR with $\mathcal{L}^G_2$. To further verify the effectiveness of the neighbor list training method, we remove $\mathcal{L}^G_2$ in Eq. \ref{equation:loss_G}, which means that PDU will lack direct guidance from NLGR-E.
\end{itemize}

The result is shown in Table \ref{tab:tab3}. From the experimental results, we have the following findings: i) The HR of NLGR w/o $r$ drops the most (8\%/17.2\%), indicating that neighbor list training is the most important part of NLGR. ii) The HR of NLGR w/ AG dropped significantly (2.2\%/4.8\%), indicating that the sampling-based non-autoregressive generation method in NLGR-G can significantly improve the generation effect. iii) The HR of NLGR w/o $\mathcal{L}^G_2$ also decreases (1.1\%/3.2\%), indicating that auxiliary loss can enhance the generation ability of NLGR-G.
\begin{table}[h]
\caption{HR of different methods on Meituan}
\label{tab:tab3}
\begin{tabular}{ccc}
\hline
\textbf{method}             & \textbf{HR@10\%}  & \textbf{HR@1\%}     \\ \hline
NLGR                        &  0.8369           &    0.7523           \\
NLGR w/o $r$                &  0.7562           &    0.5809           \\  
NLGR w/ AG                  &  0.8142           &    0.7047          \\  
NLGR w/o $\mathcal{L}^G_2$  &  0.8255           &    0.7198           \\  \hline
\end{tabular}
\end{table}

\subsection{Hyperparameter Analysis}
We analyze the sensitivity of two hyperparameters: $\alpha$ and $\beta$, corresponding to the generation process and training process of NLGR. Among them, $\alpha$ is the weight of NLGR-G loss in Eq. \ref{equation:loss_G}, and $\beta$ is the sampling ratio at each position when constructing the neighbor list $L^*$. By default $\beta=1$ means that each position is sampled $1$ time. The result is shown in Table \ref{tab:tab4}, showing the same trend on the public dataset and industrial dataset and we have the following findings:
\begin{enumerate}[leftmargin=*]
\item [i)] Hyperparameter $\alpha$ significantly affects the generator’s HR@10\% metric. When $\alpha=0$, NLGR is equivalent to the method of Group III, which means the generator has no full sight. As $\alpha$ increases, HR@10\% first increases and then decreases.

\item [ii)] We tested several values for $\beta$. When $\beta<1$, we randomly select $\beta m$ positions in $m$ positions to construct rewards. When $\beta>1$, we construct $\beta$ neighbor lists at each position. Increasing $\beta$ within a certain range can quickly improve the HR@10\% performance. As $\beta$ continues to increase, HR@10\% remains stable but increases offline training time. The results show that counterfactual rewards considering all positions are important.
\end{enumerate}

\begin{table}[h]
\caption{ HR@10\% result of ablation experiment on NLGR}
\label{tab:tab4}
\begin{tabular}{llccccc}
\hline
& $\alpha$=0 & $\alpha$=0.01 & $\alpha$=0.2 & $\alpha$=0.5 & $\alpha$=1.0  \\
\hline
Meituan     &  0.7562   &  0.8274   &  \textbf{0.8369}   &  0.8301    &   0.8295   \\
Taobao Ad   &  0.2192   &  0.3530   &  \textbf{0.4091}   &  0.3973    &   0.3912   \\
\hline
   & $\beta$=0.1 & $\beta$=0.5 & $\beta$=1 & $\beta$=2 & $\beta$=5 \\
\hline
Meituan     &  0.7763   &  0.8031   &  \textbf{0.8369}   &  0.8369    &   0.8367   \\
Taobao Ad   &  0.2814   &  0.3566   &  \textbf{0.4091}   &  0.4091    &   0.4091   \\
\hline
\end{tabular}
\end{table}

\subsection{Online A/B test}
\begin{figure}[h]
\centering
\includegraphics[width=0.45\textwidth]{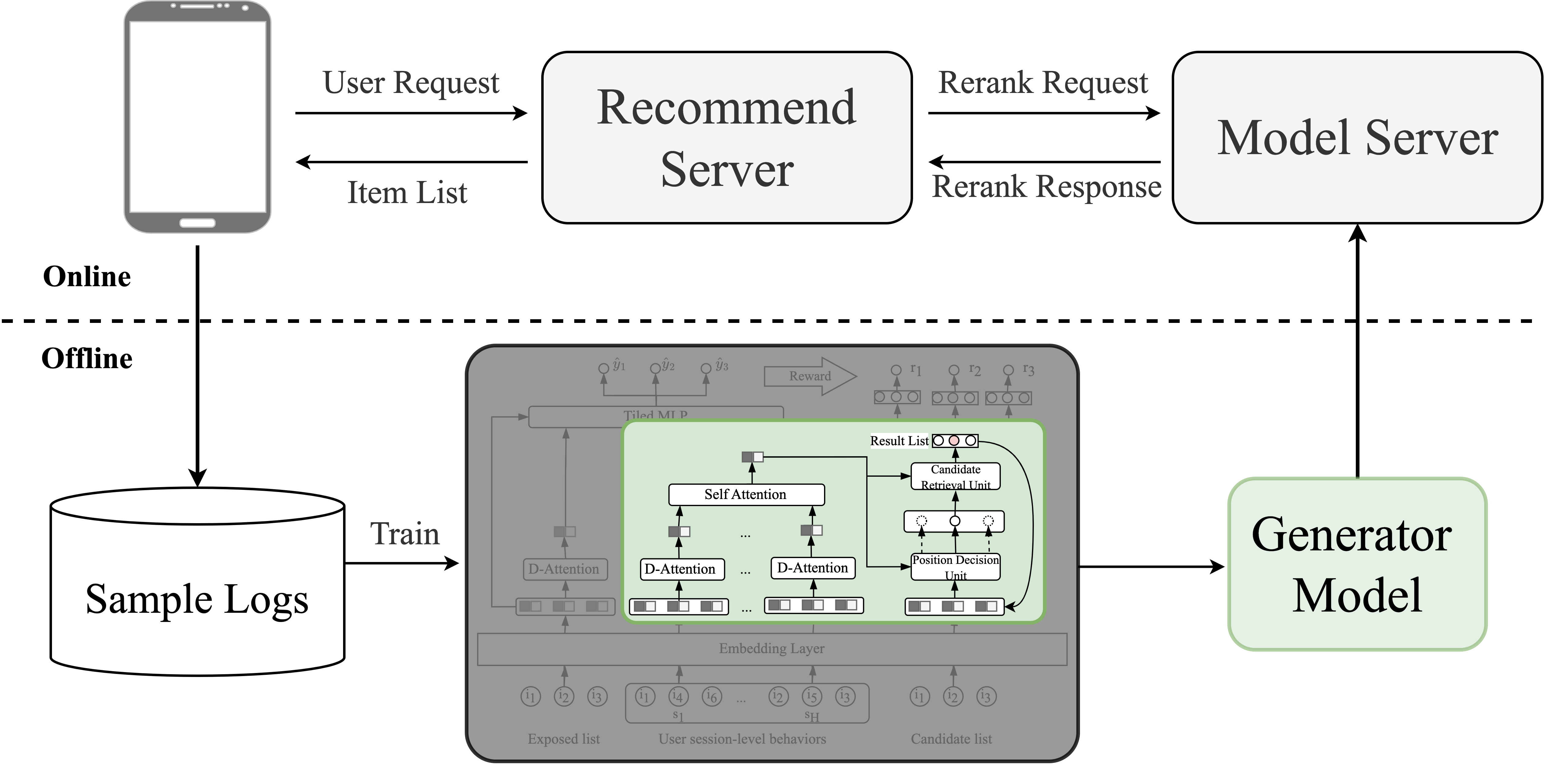}
\caption{Architecture of the online deployment with NLGR.}
\label{fig: online}
\end{figure}

We deployed NLGR in Meituan App, where Figure \ref{fig: online} shows the online serving system architecture. It is worth noting that although we involve the evaluator guiding the generator multiple times during offline training, we only need to use the generator when serving online. In this way, we ensure that its model complexity is comparable to the online model without adding additional calculations to the online service.

We conducted online experiments in an A/B test framework over five weeks from Dec.2023 to Jan.2024. Table \ref{tab:online} shows the online performance of NLGR. Compared to the baseline model (a variant of  PRM), NLGR has increased the CTR by 3.25\% and the GMV by 3.07\%. 
Moreover, since only NLGR-G is deployed online during online inference, the time-out rate online has no increase, which is acceptable to the recommendation system. Note that NLGR-E is not deployed online which is only utilized for offline guidance of NLGR-G.
Now, NLGR has been deployed in the Meituan food delivery platform and serves hundreds of millions of users.

\begin{table}[h]
\caption{Online A/B test result}
\label{tab:online}
\begin{tabular}{ccccc}
\hline
\textbf{method}        & \textbf{CTR} & \textbf{GMV}  & \textbf{Cost (ms)} & \textbf{Time-out} \\ \hline
Baseline (PRM)      &  0.0\%          &    0.0\%       & \textbf{0.0}  & 0.0\%  \\ 
\textbf{NLGR}        &  \textbf{3.25\%}& \textbf{3.07\%}& 1.6  & \textbf{0.0\%}  \\
\hline
\end{tabular}
\end{table}

\section{Conclusion}

In this paper, we make the first attempt to solve the goal inconsistency problem in reranking systems. We propose a novel framework called Neighbor List Generative Reranking (NLGR), which uses the relative scores of candidate list and neighboring lists to guide the generator. Furthermore, we propose a sampling-based non-autoregressive generator that can flexibly jump from the current list to any neighbor list. Both offline experiments and online A/B tests show that NLGR significantly outperformed other existing reranking baselines, and we have deployed NLGR on the Meituan food delivery platform.

\vspace{2em}


\bibliographystyle{ACM-Reference-Format}
\bibliography{main}

\end{document}